\documentclass[10pt]{iopart}
\usepackage{graphicx,epsf}
\begin{document}

\title[Reentrant Disordering of Colloidal Molecular Crystals...]{ 
Reentrant Disordering of Colloidal Molecular Crystals on 2D Periodic 
Substrates} 
\author{M Mikulis$^{\dag,\ddag}$, C J Olson Reichhardt$^\dag$,
C Reichhardt$^{\dag}$, R T Scalettar$^{\ddag}$ and G T Zim{\' a}nyi$^{\ddag}$} 
\address{ $^\dag$ 
Center for Nonlinear Studies and Theoretical Division, 
Los Alamos National Laboratory, Los Alamos, New Mexico 87545. \\
$^\ddag$ Department of Physics, University of California, Davis,
California 95616. }

\begin{abstract}
We study colloidal ordering and disordering on two-dimensional 
periodic substrates where the number of colloids per substrate 
minima is two or three. 
The colloids form dimer or trimer states with orientational ordering,
referred to as colloidal molecular crystals. 
At a fixed temperature such that, in the absence of a substrate,
the colloids are in a triangular floating solid state, upon 
increasing the substrate
strength we find a transition to an ordered colloidal molecular crystal 
state, {\it followed} by a transition to a disordered state where the 
colloids still form dimers or trimers but the orientational order is lost. 
These results are in agreement with recent experiments. 
\end{abstract}
\submitto{\JPCM}

\maketitle

\vskip2pc
\nosections
Colloidal particle assemblies in two dimensions (2D) are an 
ideal system to study ordering and melting, since quantities 
such as diffusion, dislocation dynamics, and local ordering can be 
directly observed, which is typically not the case for atomic and molecular
systems \cite{Murry}.  
Repulsively interacting colloids 
on a smooth substrate 
in 2D form triangular arrangements at 
high density or low temperatures, with hexatic or
liquid states at lower densities \cite{Murry}. 
Optical techniques such 
as interfering laser arrays have created one-dimensional (1D)
line-like potentials which attract the colloids.
Experimental studies of periodic 1D potentials find a
remarkable laser-induced freezing transition, where for zero substrate
strength the colloids
form a disordered liquid, but with increasing substrate or laser strength 
there is a transition to a frozen ordered state \cite{Ackerson}.   
This laser-induced freezing 
was studied with density-functional theory \cite{Theory}
and simulations \cite{Sood}. These theoretical works also found that for
increased substrate strength, the frozen state can show a reentrant 
disordering transition to a 1D modulated liquid. This reentrance was similar
to that subsequently observed in experiments \cite{Wei,Bruner}.
Further theoretical
studies have mapped additional properties of this system \cite{Frey},
and some of these predictions have also been confirmed in 
experiment \cite{Review}.     
The reentrant disordering occurs because, as the substrate strength
is gradually increased, the system becomes
effectively 1D.  The fluctuations are 
enhanced for lower dimensionality and can lead to 
the formation of a 1D liquid. 

More recently, colloidal assemblies interacting with 2D  
periodic substrates have been investigated 
\cite{Reichhardt,Bechinger,Mangold,Curtis}. Simulations of colloids on 
2D periodic substrates have considered the case where the
number of colloids is an integer multiple of the number of 
potential minima \cite{Reichhardt}. In this
case, multiple colloids sit in each potential minimum and form dimer, trimer,
or higher states depending on the filling. These colloidal states
can have an additional orientational ordering and have been termed colloidal 
molecular crystals (CMCs) in analogy with 
molecules forming crystalline states
with orientational ordering of the molecules. 
Three phases appear in the simulations:
a high temperature liquid regime, a low temperature orientationally
ordered CMC, and an intermediate disordered CMC phase
where the orientational order is lost but the colloids remain
trapped in each minimum, rotating freely but not diffusing.
Experiments on 2D periodic optical traps with three colloids
per potential minimum
confirmed the existence of the three phases \cite{Bechinger}. 
In these experiments,
the temperature and density of the system were fixed and
the intensity of the traps was increased by tuning the laser power. Here,
as a function of increasing substrate strength, the initial state was
a liquid, followed by a transition to an orientationally ordered CMC and,
at higher strengths, a transition to a disordered CMC. 
This reentrant disordering is similar to
that found for 1D periodic modulated 
substrates. In previous simulations of the CMC \cite{Reichhardt},
conclusive evidence for this reentrance was not presented. 

In this work we measure the diffusion of colloids from their traps and
also the degree of rotational order between 
molecules in adjacent wells.  We then numerically map the phase diagram
as a function of substrate strength.
We observe four phases, floating (triangular) solid, rotationally ordered
and disordered colloidal molecular crystal, and liquid.
The transition to the liquid state rises monotonically 
in temperature with the 
substrate potential strength, but 
the transition from the ordered to the disordered CMC phase 
first increases and then decreases in temperature with 
increasing substrate strength.  This leads to a striking 
feature of the phase diagram: clear evidence for a reentrant 
disordered CMC as a function of substrate strength.

We perform a Langevin simulation of a 2D system 
with periodic boundary conditions in the $x$ and $y$ directions,
as in previous work \cite{Reichhardt}.
There are $N$ colloids interacting with a 2D periodic substrate which has
$M$ minima. We focus on the two cases $N/M = 2$ and $3$.  
The overdamped equation of motion for 
a single  colloid $i$ is  
\begin{equation}
\frac{d {\bf r}_{i}}{dt} = {\bf f}_{i} + {\bf f}_{s} + {\bf f}_{T} \ ,
\end{equation}
where ${\bf f}_{i} = \sum_{j \neq i}^{N}\nabla_i V(r_{ij})$ 
is the interaction force from the other colloids, which
we take to be a Yukawa or screened Coulomb form, 
$V(r_{ij}) = (Q^2/|{\bf r}_{i} - {\bf r}_{j}|)
\exp(-\kappa|{\bf r}_{i} - {\bf r}_{j}|)$. 
Here  $Q$ is the charge of the particles, which we set to 1.0, $1/\kappa$ is 
the screening length, and ${\bf r}_{i}$ is the position of
particle $i$.
We consider a square substrate with a lattice constant $a$ with a 
force ${\bf f}_{s} = A\sin(2\pi x/a){\bf {\hat x}} + 
A\sin(2\pi y/a){\bf {\hat y}}$. In this work we keep $a$ fixed and 
vary $A$. The screening length is fixed at $1.5a$. 
The thermal force $f_T$ comes from Langevin kicks with the properties
$<f_T>=0$ and $<f_T^i(t)f_T^j(t^{\prime})>=2k_BT\delta_{ij}\delta(t-t^\prime)$.
The initial colloidal positions are obtained by annealing from a high 
temperature and gradually cooling to zero.  

\begin{figure}
\epsfxsize=0.8\textwidth
\epsfbox{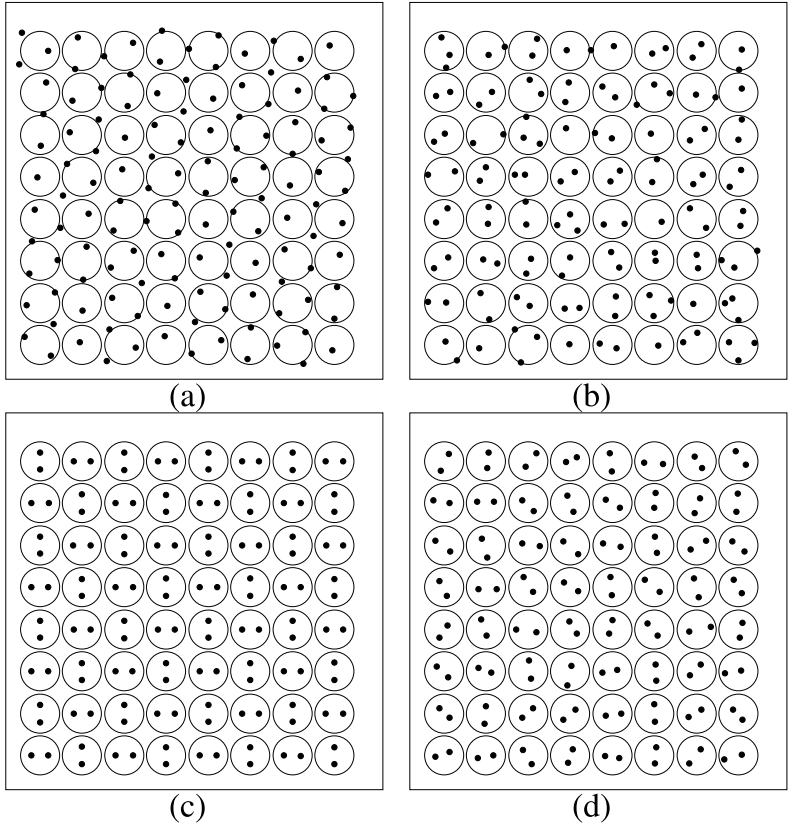}
\caption{The colloidal configurations for a square periodic substrate in 2D.  
The circles indicate the location of the potential wells.
(a) For a sufficiently weak substrate potential, at low temperatures, 
(here $A=0.0, T=0.0)$
the colloids form a floating triangular lattice, which has some mild 
distortion(see text).      
(b) At high enough temperature, 
for any substrate potential, 
(here $A=1.0, T=5.0)$
the colloids form a liquid.  (c) For all but the 
smallest potentials, at low temperatures, 
(here $A=1.0, T=0.0)$
the colloids form a rotationally 
ordered solid.  (d) For large enough substrate potentials, 
an intermediate 
(here $A=1.0, T=2.3)$
disordered CMC phase exists between the rotationally ordered solid phase and 
the liquid phase.}    
\end{figure}

We first concentrate on the case $N/M = 2$. As a function of 
$T$ and $A$, we find the four phases illustrated in figure 1. 
For very low $A$ and $T$, we find a floating triangular lattice
as shown in figure 1(a), where the substrate is weak enough that the 
elastic interaction of the colloids dominates and they form a triangular
lattice. At high enough $T$, for all $A$ we find a modulated liquid 
where the colloids are disordered and have a large diffusion constant, 
as shown in figure 1(b). 
For low $T$ and 
large enough $A$, the orientationally ordered CMC of figure 1(c) forms.
Each potential minima captures two colloids and the colloid
dimer orientation alternates from vertical to 
horizontal.
For larger $A$ and $T$, a disordered CMC phase occurs in which the
colloids are still fixed to the potential minima but
long range orientational ordering is lost, as seen in figure 1(d).   

\begin{figure}
\center{
\epsfxsize=0.8\textwidth
\epsfbox{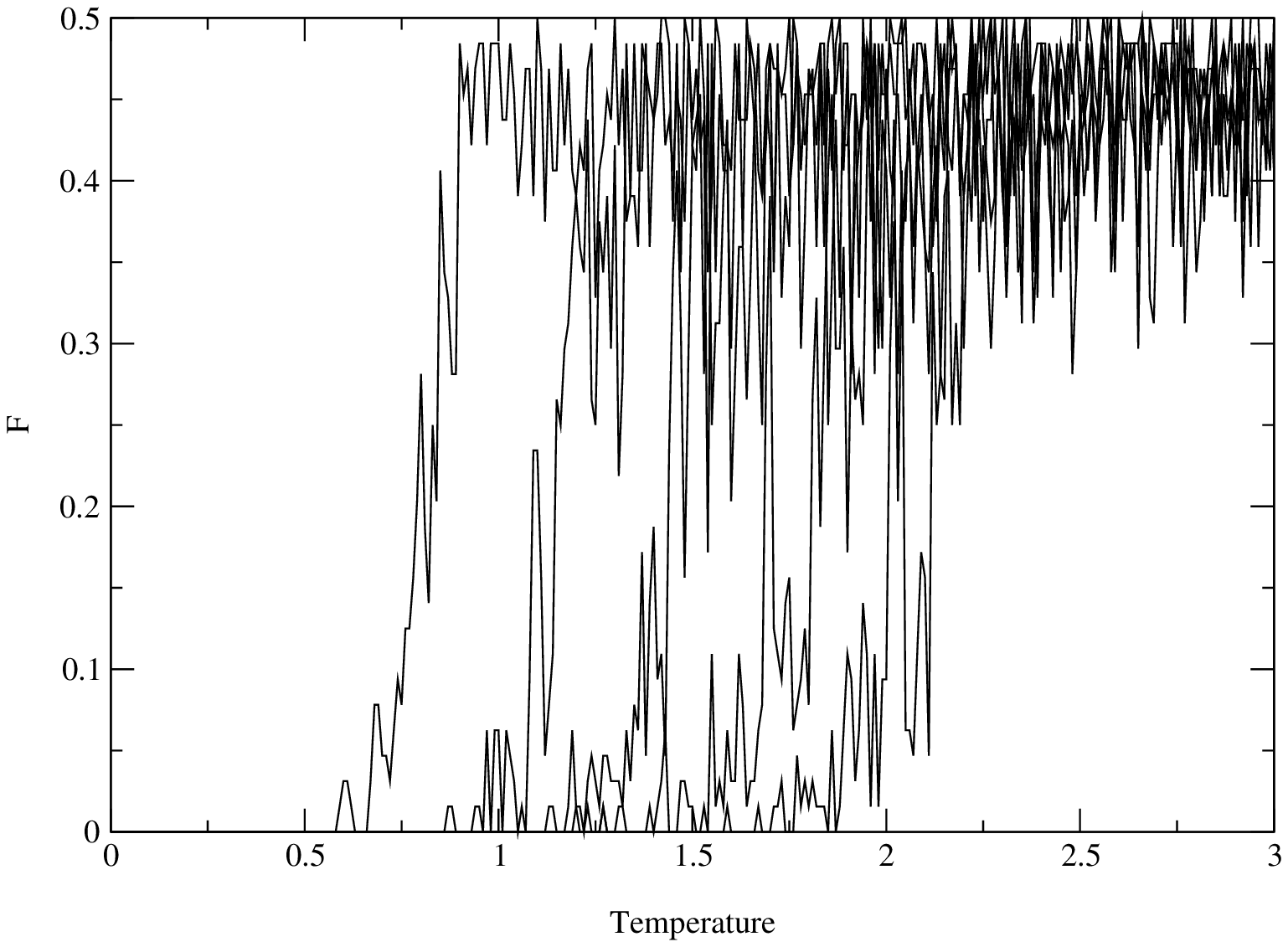}}
\caption{Rotational disorder fraction vs temperature, for (from left
to right) $A=3.0$, 2.0, 1.50, 0.25, and 0.50.  The transition line from 
ordered CMC to disordered CMC on the phase diagram in figure 4 is 
defined as the temperature 
at which $F=0.4$.  The 
transition temperature first increases from $A=0.25$
to $A=0.50$, but then decreases for larger A.}
\end{figure}

We use various measures to identify the different phases in the $T$ vs $A$ 
phase diagram. 
In order to find the ordered to disordered CMC transition, 
we measure the rotational disorder fraction $F$ of the particles.  To 
determine $F$, we define each dimer to be horizontally
or vertically oriented depending on whether the $x$ or $y$ component
of the vector joining the two colloids is larger.  $F$ is then
defined as the fraction of the dimers 
which differ 
from the ordered CMC configuration of alternating vertical and 
horizontal.  In figure 2 we plot $F$ vs $T$ for different values of $A$.

\begin{figure}
\center{
\epsfxsize=0.8\textwidth
\epsfbox{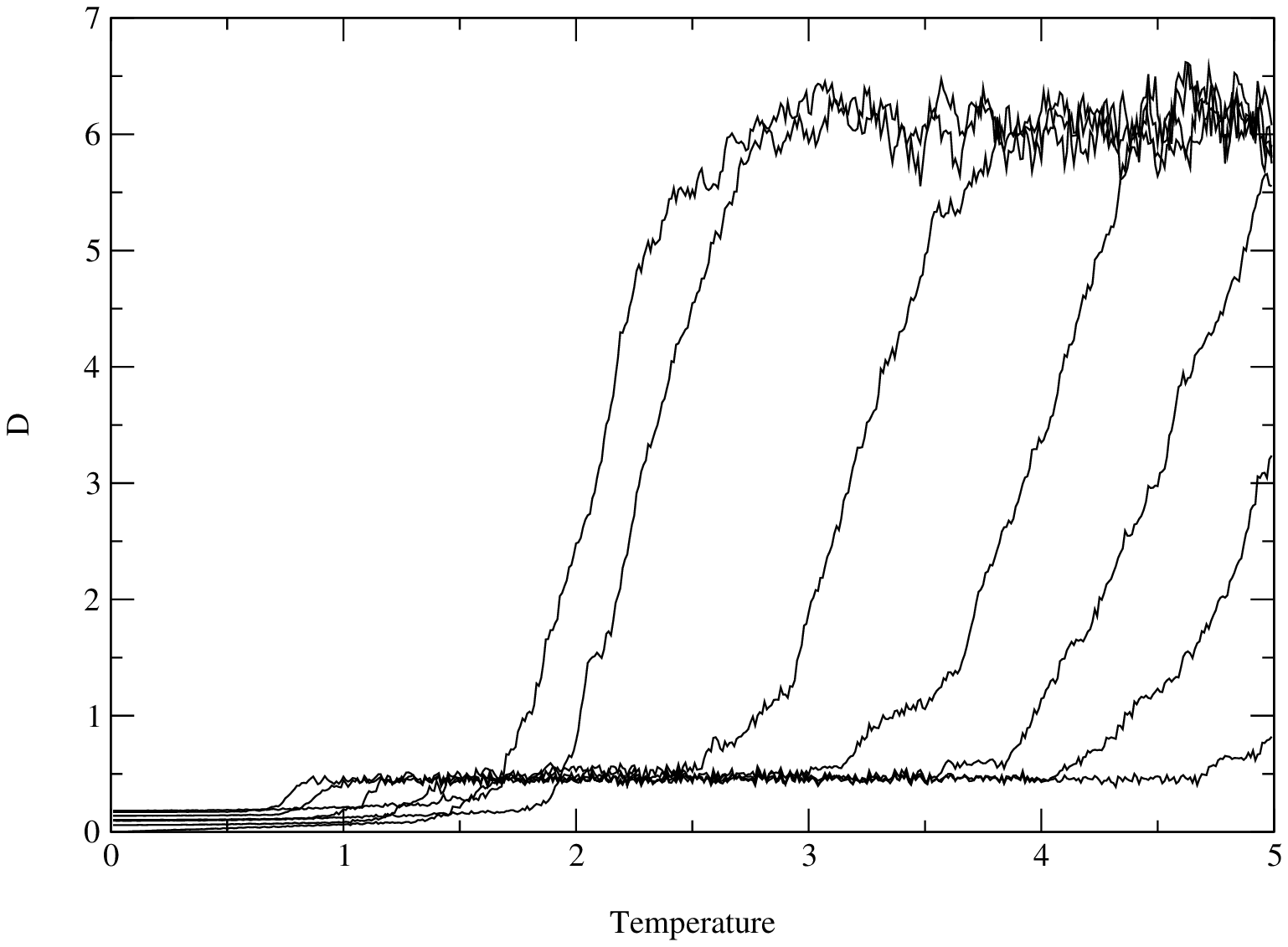}}
\caption{Diffusion vs temperature for 
(from left to right) $A=0.25$, 0.5, 1.0, 1.5, 2.0, 2.5, and 3.0.
The melting temperature is defined as the point at which $D$ first increases 
sharply, when the particles are no longer bound in the potential wells.  
}
\end{figure}

To distinguish between the CMC phases and the liquid phase, we 
measure the diffusion $D$ of the particles
by calculating the average of the square of the distance
traveled by the particles from their initial 
positions after a large, fixed number of
time steps.   In figure 3, we plot $D$ as a function of temperature
$T$ for different potential depths $A$.  The temperature at which the 
colloids escape their wells rises monotonically with $A$.
For high values of $A$, the colloid 
dimers are no longer in a rotationally ordered configuration before they leave 
the wells, showing the existence of the disordered CMC.
We measured the transitions for several
different initial starting configurations of the colloids and
found that the results were unchanged.

\begin{figure}
\center{
\epsfxsize=0.8\textwidth
\epsfbox{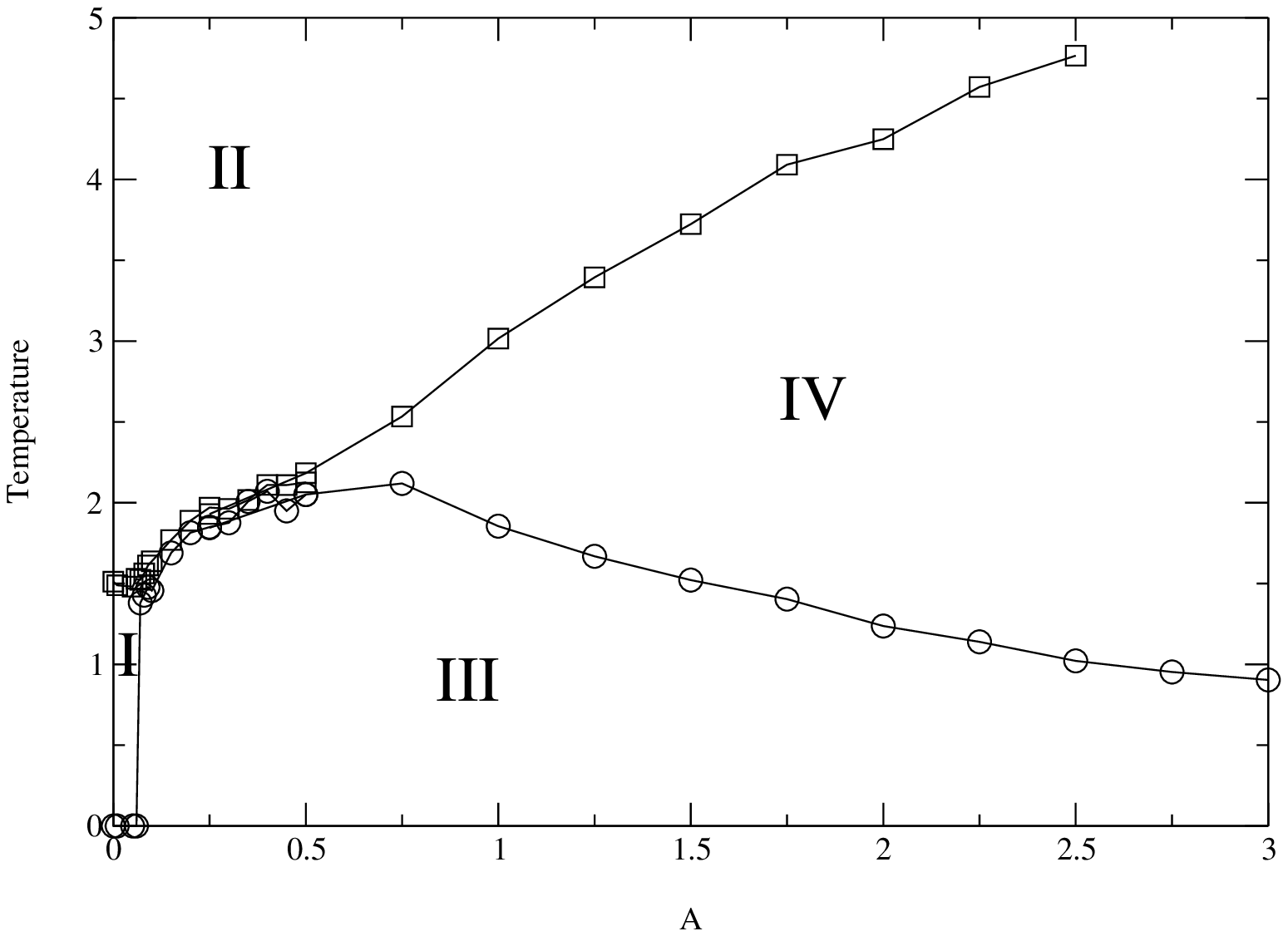}}
\caption{Phase diagram of temperature vs potential strength $A$.  
In region {\bf I} the
colloids form a floating solid of a triangular 
lattice. 
In region {\bf II} the
colloids form a liquid.  In region {\bf III} the
colloids are in a rotationally ordered solid state.  
Finally,
in region {\bf IV}
the colloids 
are in a solid state which has no rotational ordering.
Circles indicate the transition to the ordered CMC as determined
by the rotational disorder fraction $F$.  Squares indicate the 
melting transition measured from the diffusion $D$.}
\end{figure}

In figure 4 we plot the transition lines from the ordered CMC to the liquid,
floating solid, and disordered CMC states, obtained using 
the measures illustrated in figure 2 and figure 3.   
Here, the range of temperature over which the ordered
CMC phase exists increases in extent as $A$ increases from zero,
reaching a maximum width at $A = 0.75$. As $A$ is further
increased above $A=0.75$,
the width in $T$ of the ordered CMC phase
decreases.  A transition to a disordered CMC
for $A>0.5$ replaces the transition to the liquid phase for $A<0.5$.
This decrease shows a slight nonlinearity. 
The transition line between the disordered CMC and the 
liquid moves to higher temperature roughly linearly as $A$ 
increases, as expected, since it represents a first-order phase transition.
Figure 4 shows a clear disordering 
reentrance. For a fixed	
$T = 1.80$, the colloids are in a liquid state at $A = 0$. 
A transition to the ordered CMC state occurs at 
$A \approx 0.19$, while a transition to the 
disordered CMC state occurs at 
$A \approx 1.07$.  This sequence of transitions 
for increasing substrate strength is exactly what is seen in experiment
\cite{Bechinger}. 

The initial increase in the disordering temperature
with substrate strength occurs due to the fact that 
some finite substrate strength is required to confine the colloids,
which are either in a floating solid or liquid state in the absence 
of a substrate.  When the potential is just strong enough to 
trap the colloids, the size of each dimer (the distance between the
two colloids composing the dimer) is at its maximum,
and the colloids are close to jumping out of the confining well. 
At a somewhat higher temperature, 
the colloids begin to diffuse; thus, as the potential
strength increases, the temperature at which diffusion begins
also increases. We note that the cause of the
orientational ordering of the dimers is
their effective quadrupole moment. The energy of interaction between these
moments is minimized when the 
dimers are oriented as in figure 1(c).  
The strength of the quadrupole moment is 
proportional to the square of the distance
between the two colloids in a single potential minimum. As $A$ is increased,
this distance decreases. If we approximate the well confining 
the colloids as a parabola, then the distance between the colloids
in the dimer 
decreases by $\delta r \propto A$. 
As the strength of the quadrupole moment 
drops, thermal effects dominate the orientational ordering 
and the colloidal dimer begins to rotate, 
destroying the orientational ordering. 

\begin{figure}
\center{
\epsfxsize=0.8\textwidth
\epsfbox{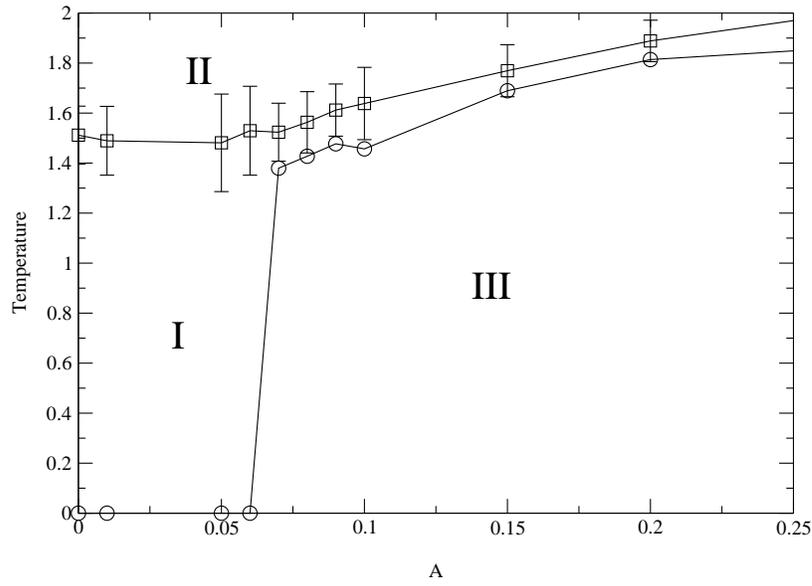}}
\caption{Phase diagram of temperature vs A at small A, highlighting 
the floating solid region.}
\end{figure}

\begin{figure}
\center{
\epsfxsize=0.8\textwidth
\epsfbox{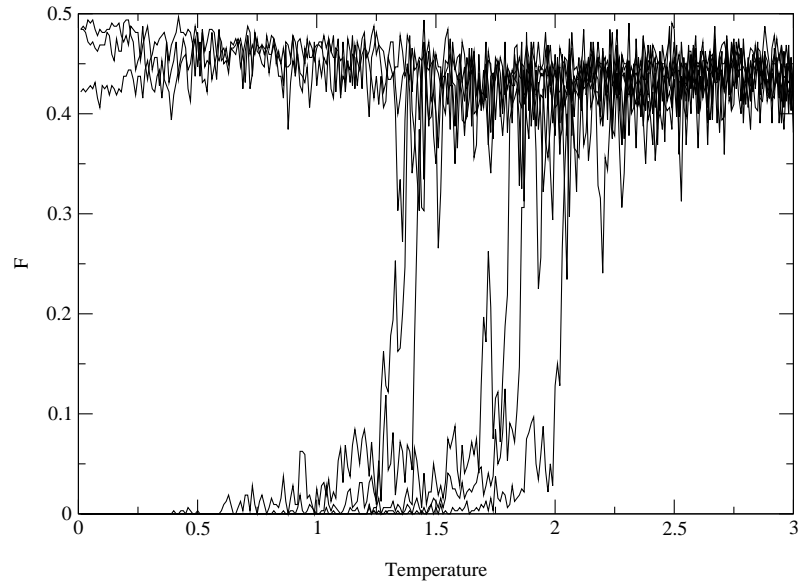}}
\caption{Rotational disorder fraction $F$ vs temperature for
(from left to right) $A=0$, 0.05, 0.06, 0.07, 0.10, 0.20, 0.30, and 0.50. 
For $A \le 0.07$, the colloids 
are initially ($T=0$) in a
floating solid state. The rotational disorder fraction exceeds 0.4 for 
these values of $A$ at all temperatures.
}
\end{figure}
 
In figure 5 we show a blowup of the phase diagram from figure 4 for the
region with small $A$. Here, we find that at low temperatures
the elastic interactions between the colloids
dominate over the substrate, and an almost triangular lattice 
forms. We call this a floating solid since the system is 
effectively frozen 
with no diffusion. 
In figure 6 we show the rotational disorder fraction plot for small 
values for A, which was used to identify the floating solid phase.  

\begin{figure}
\center{
\epsfxsize=0.8\textwidth
\epsfbox{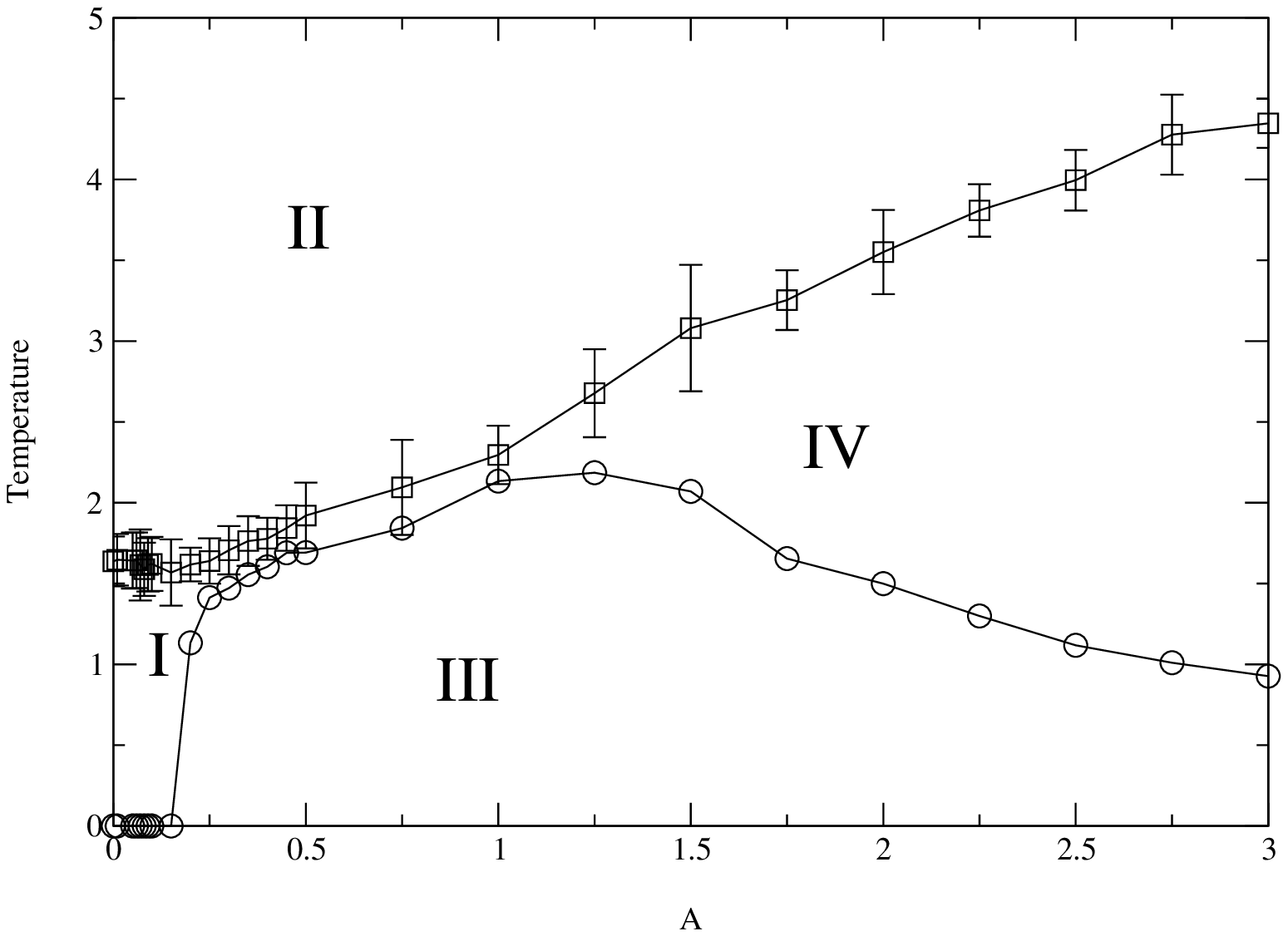}}
\caption{Phase diagram of temperature vs A for trimer states.  The 
same four phases occur as in the dimer states, namely: {\bf I} - floating 
solid, {\bf II} - liquid, {\bf III} - ordered 
CMC (colloid molecular crystal), and 
{\bf IV} - disordered CMC.}
\end{figure}

In order to show that the reentrant disordering is a general feature for
CMC states we have considered other 
integer fillings as well.
In figure 7 we plot the phase diagram 
using the same measures as in 
figure 4 for the case of $N/M = 3$ on a triangular
substrate. This is the filling fraction used in recent experiments 
\cite{Bechinger}. Here the ordered CMC is the same as found in earlier
simulations \cite{Reichhardt}. 
The phase diagram has the same basic features as in figure 4,
where $M/N=2$ and dimers form in the wells.
However, the critical temperatures are a bit lower,
as is reasonable, and while the non-monotonic (re-entrant) behavior of
the rotational melting line is still clearly evident,
the liquid at small $A$ does not extend as
deeply below the peak in the rotationally ordered phase.

In conclusion, we have shown clear numerical evidence for reentrant 
disordering of colloidal molecular crystals on 2D periodic substrates
for increasing substrate strength 
and fixed temperature, in agreement with recent experiments. 
We map out the temperature vs substrate phase diagram and show that 
four phases can occur: a high temperature liquid, a triangular
floating solid, an orientationally ordered colloidal molecular crystal state,
and a disordered colloidal molecular crystal state.

\ack
We thank C. Bechinger for useful discussions. 
This work was supported by the US DoE under Contract No. W-7405-ENG-36,
by the LANL/UC CARE and CLE programs, and by NSF-DMR-0312261.

\end{document}